\documentclass{piparticle-final}
\usepackage[utf8]{inputenc}
\usepackage{amsmath}
\usepackage{graphicx}

\usepackage{epstopdf} 

\begin{document}

\volume{7}               
\articlenumber{070011}   
\journalyear{2015}       
\editor{C. A. Condat, G. J. Sibona}   
\received{20 November 2014}     
\accepted{24 June 2015}   
\runningauthor{E. Hern\'andez-Lemus \itshape{et al.}}  
\doi{070011}         

\title{The role of master regulators in gene regulatory networks}

\author{E. Hern\'andez-Lemus,\cite{inst1}\thanks{ehernandez@inmegen.gob.mx} \hspace{0.3em}
K. Baca-L\'opez,\cite{inst1}
R. Lemus,\cite{inst1}
R. Garc\'ia-Herrera\cite{inst1}}

\pipabstract{
Gene regulatory networks present a wide variety of dynamical responses to intrinsic and extrinsic perturbations. Arguably, one of the most important of such coordinated responses is the one of amplification cascades, in which activation of a few key-responsive transcription factors (termed master regulators, MRs) lead to a large series of transcriptional activation events. This is so since master regulators are transcription factors controlling the expression of other transcription factor molecules and so on. MRs hold a central position related to transcriptional dynamics and control of gene regulatory networks and are often involved in complex feedback and feedforward loops inducing non-trivial dynamics.
Recent studies have pointed out to the myocyte enhancing factor 2C (MEF2C, also known
as MADS box transcription enhancer factor 2, polypeptide C) as being one of such master
regulators involved in the pathogenesis of primary breast cancer. In this work, we perform an
integrative genomic analysis of the transcriptional regulation activity of MEF2C and its target
genes to evaluate to what extent are these molecules inducing collective responses leading to gene expression deregulation and carcinogenesis. We also analyzed a number of induced dynamic
responses, in particular those associated with transcriptional bursts, and nonlinear cascading
to evaluate the influence they may have in malignant phenotypes and cancer.
}

\maketitle

\blfootnote{
\begin{theaffiliation}{99}
   \institution{inst1} Computational Systems Biology and Integrative Genomics Lab, Computational Genomics
Department, National Institute of Genomic Medicine, Perif\'erico Sur 4809, Col. Arenal Tepepan, M\'exico 14610, D.F. M\'exico.
\end{theaffiliation}
}

\section{Introduction: Transcriptional master regulators}

Phenotypic conditions in living cells are largely determined by the interplay of a multitude of molecules; in particular, genes and their protein products. The coordinated behavior of such a large number of players is often represented by means of a gene regulatory network (GRN). In a GRN, regulatory processes between genes, transcription factors and other molecular components are represented by nodes and links. One common way of inferring this gene regulatory networks is by probabilistic analysis of whole genome gene expression data \cite{ehlphysa,aracne}.\\

Specific, context-dependent analysis of regulatory activity of particular cellular phenotypes (say tumor cells) may also be performed with the aid of transcriptional interaction networks. Commonly, such GRNs present a complex topology, often compliant with a scale-free hierarchic nature, in which a relatively small number of key players dominate the function and dynamics of the network. Some of these key players in GRNs are transcription factors often known as \emph{master regulators} (MRs). MRs are deemed responsible for the control of the whole regulatory program for cells under the associated phenotype \cite{basso,muna}. Master regulators may, indeed, act over rather generalistic cellular processes \cite{mtor}, but also on specific cellular phenotypes \cite{muna,pax5,gcn4p}. \\

For instance, it is known that the mTOR molecule is active in concerting signals regulating control growth, metabolism, and longevity. Malfunction of mTOR complexes has been associated with developmental abnormalities, autoimmune diseases and cancer \cite{mtor}. The main role of mTOR seems to be the regulation of protein synthesis. Detailed mechanisms remain unknown, but ribosome profiling seems to point out to translational regulation and transcriptional activation activity. Due to the multiplicity of mTOR signaling interactions, this molecule acts as a master regulator on a variety of phenotypes. More specific master regulatory activity may be exemplified by cases such as the one of VASH1 that has been identified as a master-regulator of endothelial cell apoptosis \cite{muna}; PAX5 is known to be a master regulator of B-cell development also involved in neoplastic processes in leukemogenesis \cite{pax5} and the yeast protein Gcn4Pp (that contains a conserved domain cd12193 present in human JUN proto-oncogene) that is triggered by starvation and stress signals \cite{gcn4p} and is an MR in the phenotypic response to such stimuli. Due to the complex mechanisms behind transcriptional regulation in eukaryotes, identification of MRs is mostly based on the (inferred or empirical) relationships between them and their downstream RNA targets in the GRN.\\

In brief, MRs are transcription factor genes that are located \emph{upstream} in the genomic regulation programme, hence they possess a high hierarchy in the GRN. They are considered to be important players behind the presence of (some) amplification cascades in transcriptional regulatory networks, and it has been hypothesized that they may coordinate the dynamic transcriptional response and phenotype (in the case of eukaryotes) of the cells.\\ 

As it may be evident, MRs may have a big impact on cancer-related phenotypes. This is so since under genome instability conditions, the uncontrolled synthesis of these molecules may give rise to large amplification of transcriptional cascades. In Ref. \cite{karolplos},  the role that some molecules (in particular, MEF2C) may have in processes involving metabolic deregulation and MR activity at the onset of primary breast cancer was studied. The approach followed there involved the inference of GRNs centered in a number of molecules considered to be candidate MRs associated with the breast cancer phenotypes at early stages (i.e., primary tumors). As it can be seen there, MEF2C resulted a quite promissory molecule due to the large number of (probabilistically inferred) targets it possesses, but also due to the main biological processes spanned by its targets.\\

\subsection{A \emph{what if?} scenario}

Now, let us resort for a moment to a hypothetical scenario: Imagine you have a  eukaryotic cell with \emph{deregulated metabolism} (e.g., large local free energy fluctuations) and a gene with transcription factor activity that has a low activation energy threshold (i.e., a relatively low absolute value for the free energy of formation). Now imagine that this gene is located (within the regulatory network) \emph{close} to energy transduction pathways and that it also possesses a high hierarchy on the transcriptional regulatory network as well as a relatively large relaxation time.\\

To put it more clearly, the fact that a gene has low activation energies means that the amount of energy needed to activate its cellular biosynthesis is minimum \cite{activa1, activa2}, thus making this molecule more prone to be produced by large free energy fluctuations. The probability to have such large energy fluctuations may be increased under abnormal metabolism conditions \cite{epste, karolplos, tenant} that may enable events leading to transcriptional cascades.\\

In such scenario, large local free energy fluctuations may \emph{randomly} activate the transcription of such a gene that in turn may be able to activate long ranged transcriptional cascading before decaying, thus affecting to a large degree the whole transcriptional regulation programme of such cell. Indeed, such a gene may be acting as a transcriptional master regulator over that specific cell condition (phenotype).

\section{MEF2C as a master regulator}

\begin{figure}[ht]
\begin{center}
\includegraphics[width=0.8\columnwidth]{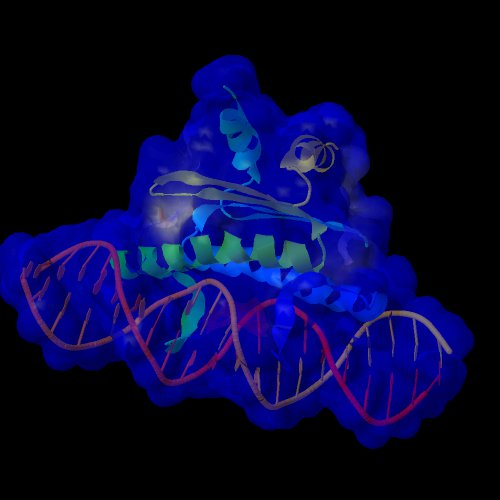}
\end{center}
\caption{Poisson-Boltzmann visualization of the MEF2C transcription factor protein showing the action of both, mef2 and MADS-box DNA-binding domains.}
\label{mef2cst}
\end{figure}

In Ref. \cite{karolplos}, we discussed the evidence that may point out to the MEF2C molecule as a candidate master regulator for the transcriptional regulator of human cells under the primary breast cancer phenotype.  Regarding this molecule, we know the following: MEF2C is a transcription factor gene located (in humans) in 5q14.3 on the minus strand. This gene is 200,723 bp long and it encodes a 473-aminoacid protein weighting 51.221 kDa. MEF2C is a member of the Mef2 family that by means of controlling gene expression (MEF2 molecules are commonly acting as activator transcription factors) is able to regulate cellular differentiation and development \cite{mef2}. MEF2 members are highly versatile regulators since they contain both MEF2 and MADS-box DNA binding domains (see Fig. \ref{mef2cst}). The MADS-box serves as the minimal DNA-binding domain, however an adjacent 29-amino acid extension called the Mef2 domain is required for high affinity DNA-binding and dimerization, hence conferring a combinatorial DNA binding repertoire through a number of transcription factor recognition marks \cite{meff}.\\

It is also known that the MEF2C protein interacts with MAPK7 (involved in proliferation and differentiation signaling) \cite{mapk7}, EP300 (a transcription factor that regulates cell growth and cellular division) \cite{ep300}, TEAD1 (an enhancer TF that co-regulates transcription with MEF2C), as well as with a number of histone deacetylases, most notably HDAC4, HDAC7, and HDAC9 \cite{hdac,hdac2}. These protein-protein interactions, mostly with other transcription factors, enhancers or epigenomic regulators joined with their inherent binding-site transcriptional activity made MEF2C a quite functional and adaptable MR.\\

Aside from this, a non-equilibrium thermodynamics analysis of the coupling between transcriptional regulation processes and metabolic de-regulation in breast cancer cells has led to some further evidence pointing out to MEF2C as an MR that may be playing an important role in carcinogenic processes. This thermodynamic evidence has been supplemented with information given by probabilistically-inferred gene regulatory networks centered around genes coordinating the coupling of transcriptional control and metabolism. The GRN was inferred by mutual information calculations \cite{itgrn,casm} on a database of 1191 whole genome gene expression experiments in biopsy captured tissue from primary breast cancer patients and healthy controls \cite{karolplos}.\\

\begin{figure}[ht]
\begin{center}
\includegraphics[width=1.0\columnwidth]{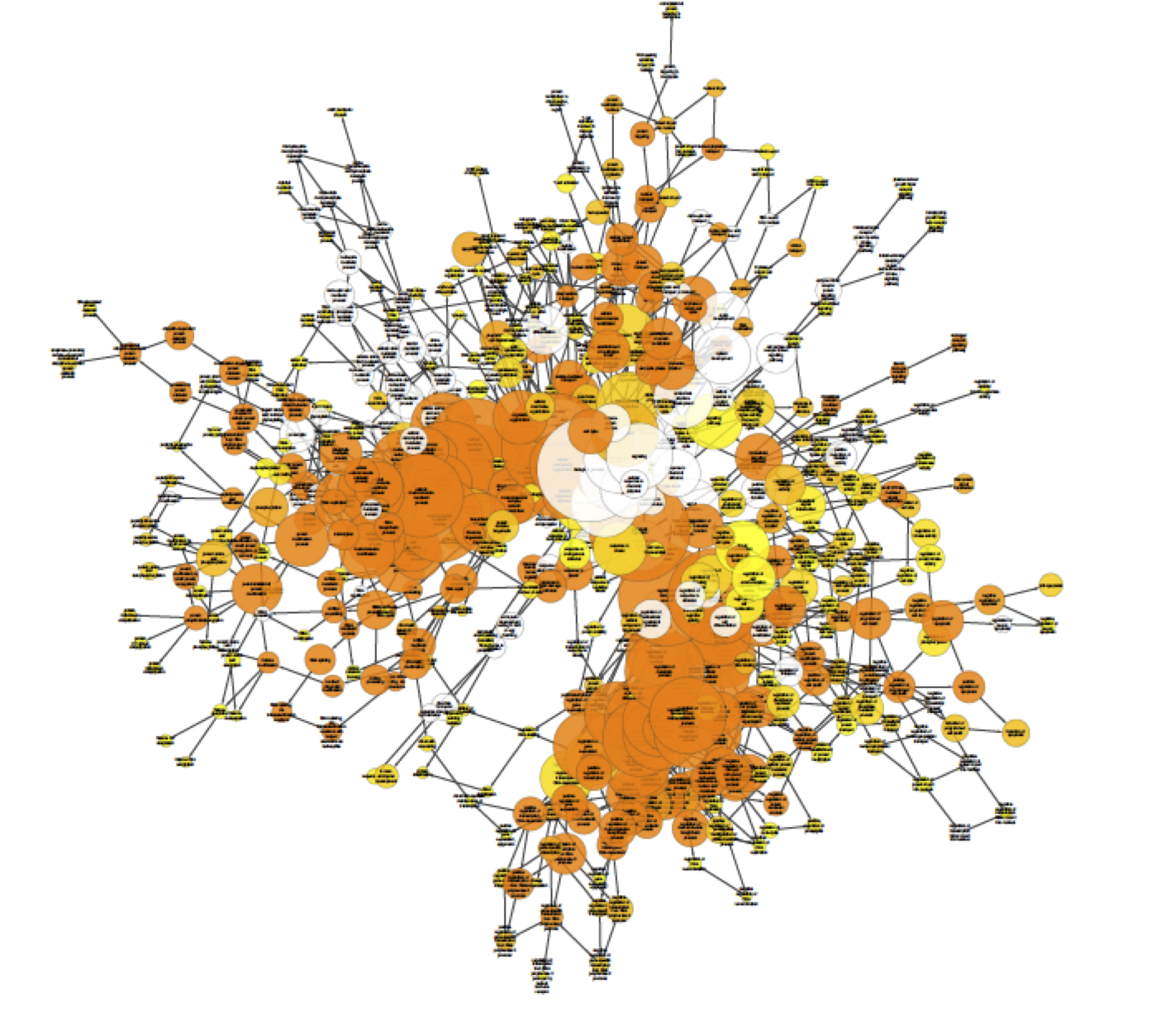}
\end{center}
\caption{Hierarchical network displaying statistical enriched Gene Ontology Biological Process entries related with MEF2C cascading, we may see that this evidence supports the hypothesis made in Ref. \cite{karolplos} regarding a non-equilibrium coupling between metabolism and transcriptional regulation.}
\label{metab}
\end{figure}

To further express that this behavior is related to the coupling between transcriptional regulation control and energy transduction pathways, let us resort to Fig. \ref{metab}. Figure \ref{metab} shows a gene ontology network containing the biological processes statistically enriched in a list of MEF2C regulated genes, differentially expressed in a 1191-sample database of whole genome gene expression experiments curated in our group \cite{karolplos}. In this figure, we may see that statistically enriched biological processes are shown as color-coded (white to red) according to a p-value calculated from a hypergeometric urn model test and corrected via the false discovery rate (FDR) measure as it is explained elsewhere \cite{karolplos}. We can see that the two major families of biological processes enriched are precisely those related with energy-release metabolic pathways and transcriptional regulation.

\subsubsection{Transcription factor binding site analysis}

In order to further validate the findings given by non-equilibrium thermodynamics and probabilistic regulatory networks, a computational analysis of databases for DNA transcription factor binding sites (TFBS) was performed. This study included a systematic TFBS analysis for MEF2C transcriptional influence by applying an algorithm (MotEvo) \cite{motevo} that incorporated -via Bayesian optimization- information additional to the sequence (physicochemical and electrostatic features, motif conservation and phylogeny, ChIP experiments, DeepCAGE sequencing, etc.). Such analysis was performed with a stringent statistical significance level  (Response values $>$ 1.5 corresponding approx. to p* $<$ 0.001) and showed that genome-wide MEF2C is able to regulate 200 genes directly.\\

The results of the TFBS analysis showed that the set of MEF2C targets includes a number of genes that participate in oncogenic processes such as MRAS, IGFBP3, CTNND1, FOXN3, FOXP4, HGMA2, MMP19, CORO1C, JAG1, ASXL1, HSPB1, MB, RBL2, ZIC2, NR2F6, BCL-2, CBX7, DNM2, MAFA, LGALS3BP, among others. Additionally, a number of MEF2C targets are in turn TFs, which enlarges the range of transcriptional influence of MEF2C (and help it to become an MR).\\

\begin{figure}[ht]
\begin{center}
\includegraphics[width=1.0\columnwidth]{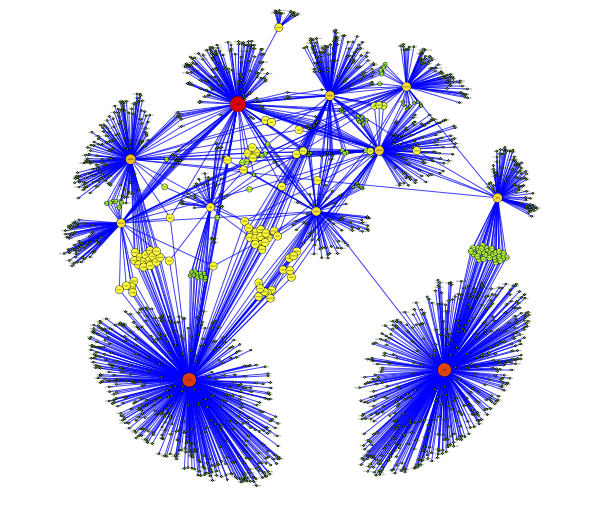}
\end{center}
\caption{Gene regulatory network including up to second order transcriptional interactions in MEF2C targets. The network is composed of 1896 genes and 2156 interactions.}
\label{mef2cgrn}
\end{figure}

In fact, second order transcriptional interactions increase the range of influence of MEF2C to a GRN composed of 1896 genes and 2156 regulatory interactions (see Fig. \ref{mef2cgrn}) that was able to further increase to 5852 genes and 18801 interactions up to third order.

\section{Dynamics of master regulator activity}

Cells sometimes present bursts or pulses of activity in their gene expression dynamic patterns. Bursting may result from a series of stochastic biochemical events and may be a source of large phenotypic heterogeneity of cell behavior and thus on cellular conditions and disease. Noise in transcriptional regulatory activity arises not only as a consequence of  randomness of biochemical processes at the molecular level due to low molecule counts, it also may arise from thermodynamic fluctuations in cellular components and system level phenomena due to cooperativity. Different levels of gene regulation may be strongly coupled. When all these elements are present, we say that the cells are undergoing transcriptional bursts (TBs). Under such dynamic scenario, protein production occurs in pulses, each due to a single promoter or transcription factor binding event. It is in these instances that the phenomena can be related to the presence of master regulators in the transcriptional networks.

\subsubsection{Bursting and synchronization}

The non-linear behavior of GRN interactions can be better understood in the light of periodic or quasi-periodic expression levels for certain groups of genes. By means of \emph{Power Spectral Density} (PSD) calculations we may gain greater insight in such dynamic behavior.  Power spectral density is useful to describe the evolution of the variance that in turn provides us with greater insight on the correlation structure of the underlying regulatory processes. The power spectral density of stochastic quasi-stationary processes can be estimated when considering non-linear time series analysis as follows.

Let us consider $\Gamma$ as a series containing a time course for intensity levels of  gene expresion of a single gene, then the associated power spectral density, $I(\omega)$ is given by:

\begin{equation}\label{psd2} I(\omega) = \frac{1}{N}\left|\sum_{t}^N \Gamma(t) \; \exp(-i \; \omega t) \right|^{2}; \; \; \, \omega \in [0, \pi] \end{equation}

Periodic behavior could be detected in a \emph{linear} model for $\Gamma$:

\begin{equation}\label{gammaper} \Gamma(t) = \beta \; \cos(\omega t + \phi) + \epsilon_i \end{equation}

$\beta$ being a positive constant (amplitude), $\omega \in [0, \pi]$, $\phi$ is a uniformly distributed \emph{phase shift} ($\phi \in (-\pi, \pi]$) and \{$\epsilon_i$\} is a sequence of uncorrelated random variables with mean $0$ and variance $\sigma^2$ independent of $\phi$ (i.e. a Gaussian noise). Under this model, periodic behavior could be traced-off by means of looking at \emph{significant} peaks in the power spectral density, either within an $\omega$-continuous process or more commonly with $\omega$ taking discrete values $\frac{2\pi k}{N}; \; \; k=0,1,2 \dots , [\frac{N}{2}]$, each of these discrete values is known as a Fourier frequency.\\

If a time course $\Gamma$ has hidden periodic components, say with a given frequency $\omega^\star$, then the power spectral density will show a peak at $\omega^\star$. If, on the other hand, $\Gamma$ is a random, aperiodic signal, then the $I(\omega)$ Vs $\omega$ plot will be a (noisy) straight line, mapping to $\beta = 0$ in the linear model as given by Eq. \ref{gammaper}. Then we may test the null hypothesis $\beta = 0$ versus the data in order to determine \emph{significance} . An early result from Fisher applies also to finite time series, the so called g-statistic \cite{fisher}:

\begin{equation}\label{fishg} g = \frac{\max_k \; I(\omega_k)}{\sum_{k=1}^{N/2} \, I(\omega_k)}\end{equation}

Values of $g$ larger than \emph{expected} lead to the rejection of the null hypothesis (i.e., random processes). The exact $g$-distribution  is given by:

\begin{eqnarray}\label{gege}
P(g > x) &=&  \alpha (1-x)^{\alpha-1} - \frac{\alpha (\alpha-1)}{2} (1-2x)^{\alpha-1} \nonumber \\
  &+& \dots + (-1)^r \frac{\alpha !}{r! (\alpha -r)!} (1-rx)^{\alpha-1}
\end{eqnarray}

$\alpha = N/2$ and $r$ is the largest integer less than $1/x$. So if the observed value of $g$ is $g^\star$, then there is a p-value $P(g > g^\star)$ to evaluate the null hypothesis.\\

\begin{figure*}[ht]
\begin{center}
\includegraphics[width=0.9\textwidth]{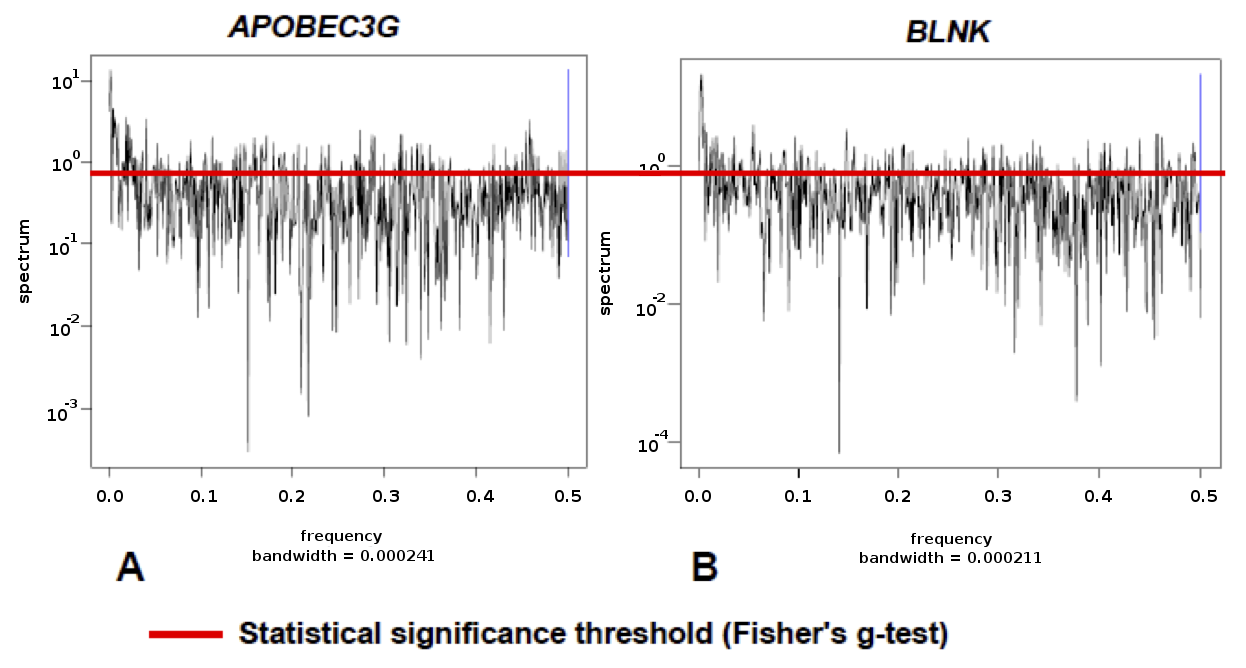}
\end{center}
\caption{Power Spectral Density plots for the time course of two MEF2C-regulated genes (APOBEC3G and BLNK) in a database of whole genome gene expression for primary breast cancer samples.}
\label{burst}
\end{figure*}

In Fig. \ref{burst}, we can see the results of the application of Fisher's significance analysis (red line) to the power spectral density profiles of two genes that are regulated by MEF2C. We can see that both genes present some significant peaks in the power spectral density profiles, which means that some quasi-periodicities are present (i.e., some frequency bands are enriched).  Further analyses have shown that the peaks for different MEF2C targets are indeed heavily correlated in some instances. This is related to a mechanism of functional synchronization. For a more detailed description of such analysis, please see Ref. \cite{rmfchaos}.

\section{Conclusions}

As a consequence of the arguments just exposed here, we have been able to reach some conclusions about the role that master regulators (in this case MEF2C) may be playing in the function and dynamics of gene regulatory networks for particular phenotypes (in this case potentially related to the onset of primary breast cancer).\\

In the first place, we may mention that MEF2C, as a transcriptional master regulator, associated with tumor phenotypes has a number of important physico-chemical features: such as having a low activation energy, and long decay times. MEF2C also presents TFBSs of two quite general classes (MADS and MEf2) which makes it a highly versatile transcription factor molecule. Following stringent TFBS analyses, MEF2C is potentially involved in the regulation of up to 200 genes directly, about 2000 at second order, and almost 6000 at third order (almost one fourth of the entire human genome) and also a number of its target genes are, in turn, transcription factors some of them with global activity. However, we must stress that MEF2C is not the only molecule responsible for the regulatory programme of such molecules. To what extent its coordinated action is able to induce the phenotype \emph{in vivo} is still something to be determined by further experimental data.\\

In relation to the dynamics of biological processes induced by MEF2C cascading, we have seen that MEF2C targets present ‘stochastic bursting’ and such bursts in gene expression activity are synchronized and long range correlated to a high degree (i.e., Almost 1/f correlations).
Dynamic synchronization and long range correlation appear to be functional biological phenomena. However, ad hoc experimental testing is still in design.\\

With regards to the biological implications of such findings (especially in the context of cancer biology), we have observed that MEF2C may be associated with tumor phenotypes (at least in primary breast cancer, according to our results). This is so because MEF2C has a number of direct targets (and also many of the indirect ones) which are molecules with known oncogenic activity. The patterns of gene expression of MEF2C targets (even for genes that are not differentially expressed) are able to induce the phenotype consistently. Also, statistical enrichment analyses, both for biological processes and biochemical pathways, showed significant hits associated with cancer related categories.\\

After this, all that one has to say is that a lot of work is still to be done to understand the complex mechanisms behind the Master Regulatory control of phenotypes (especially in disease-related scenarios). Such investigations need to be multidisciplinary in nature and must be anchored in solid mathematical foundations compliant with the tenets of the theory of complex systems, while at the same time must be heavily relying on solid biological knowledge. Comprehensive integrative analyses under the systems biology paradigm will surely hence be a must at the center of discussion on these matters.





\begin{acknowledgements}
The authors gratefully acknowledge support by grant 179431/2012 (CONACYT), as well as federal funding from the National Institute of Genomic Medicine (M\'exico).
\end{acknowledgements}


\begin{thebibliography}{99}

\bibitem{ehlphysa} E Hern\'andez-Lemus,  D Vel\'azquez-Fern\'andez, J K Estrada-Gil, I Silva-Zolezzi, M F Herrera-Hern\'andez, G Jim\'enez-S\'anchez, {\it Information theoretical methods to deconvolute genetic regulatory networks applied to thyroid neoplasms}, Physica A {\bf 388}, 5057 (2009).

\bibitem{aracne} A A Margolin, I Nemenman, K Basso, C Wiggins, G Stolovitzky, A Califano, {\it ARACNe: An algorithm for the reconstruction of gene regulatory networks in a mammalian cellular context}, BMC Bioinformatics {\bf 7} (Suppl I), S7 (2006).

\bibitem{basso} K Basso, A A Margolin, G Stolovitzky, U Klein, R Dalla-Favera, A Califano, {\it  Reverse engineering of regulatory networks in human B cells},  Nat. Genet. {\bf 37} 382 (2005).

\bibitem{muna} M Affara, D Sanders, H Araki, Y Tamada, B J Dunmore, S Humphreys, S Imoto, C Savoie, S Miyano, S Kuhara, D Jeffries, C Print, D S Charnock-Jones, {\it Vasohibin-1 is identified as a master-regulator of endothelial cell apoptosis using gene network analysis}, BMC Genomics {\bf 14}, 23 (2013).

\bibitem{mtor} R Hosking, {\it mTOR: The master regulator}, Cell {\bf 149}, 955 (2012).

\bibitem{pax5} J. Medvedovic, A Ebert, H Tagoh, M Busslinger, {\it Pax5: A master regulator of b cell development and leukemogenesis}, Adv. Immunol. {\bf 111}, 179 (2011).

\bibitem{gcn4p} A G Hinnebusch, K Natarajan, {\it Gcn4p, a master regulator of gene expression, is controlled at multiple levels by diverse signals of starvation and stress}, Eukaryot. Cell. {\bf 1}, 22 (2002).

\bibitem{activa1} S J Maerkl, S R Quake, {\it A systems approach to measuring the binding energy landscapes of transcription factors}, Science {\bf 315}, 233 (2007).

\bibitem{activa2} M Sawadogo, R G Roeder, {\it Energy requirement for specific transcription initiation by the human RNA polymerase II system}, J. Biol. Chem. {\bf 259}, 5321 (1984).

\bibitem{epste} T Epstein, L Xu, R J Gillies, R A Gatenby, {\it Separation of metabolic supply and demand: Aerobic glycolysis as a normal physiological response to fluctuating energetic demands in the membrane}, Cancer Metabolism {\bf 2}, 7 (2014).

\bibitem{karolplos} K Baca-L\'opez, A Hidalgo-Miranda, M Mayorga, N Guti\'errez-N\'ajera, E Hern\'andez-Lemus, {\it The role of master regulators in the metabolic/transcriptional coupling in breast carcinomas}, PLoS ONE {\bf 7}, e42678 (2012).

\bibitem{tenant} D A Tennant, R V Durn, E Gottlieb,  {\it Targeting metabolic transformation
for cancer therapy}, Nat. Rev. Cancer {\bf 10}, 267 (2010).

\bibitem{mef2} M J Potthoff, E N Olson, {\it MEF2: A central regulator of diverse developmental programs}, Development {\bf 134}, 4131 (2007).

\bibitem{meff} J D Molkentin, E N Olson, {\it Combinatorial control of muscle development by basic helix-loop-helix and MADS-box transcription factors},  P. Natl. Acad. Sci. USA {\bf 93}, 9366 (1996).

\bibitem{mapk7} C C Yang, O I  Ornatsky, J C McDermott, T F Cruz, C A Prody, {\it Interaction of myocyte enhancer factor 2 (MEF2) with a mitogen-activated protein kinase, ERK5/BMK1}, Nucleic Acids Res. {\bf 26}, 4771 (1998).

\bibitem{ep300} V Sartorelli, J Huang, Y Hamamori, L Kedes, {\it Molecular mechanisms of myogenic coactivation by p300: Direct interaction with the activation domain of MyoD and with the MADS box of MEF2C}, Mol. Cell. Biol. {\bf 17}, 1010 (1997).

\bibitem{hdac} A H Wang, N R Bertos, M Vezmar, N Pelletier, M Crosato, H H Heng, J Th'ng, J Han, X J Yang, {\it DAC4, a human histone deacetylase related to yeast HDA1, is a transcriptional corepressor}, Mol. Cell. Biol. {\bf 19}, 7816 (1999).

\bibitem{hdac2} A H Wang, X J  Yang, {\it Histone deacetylase 4 possesses intrinsic nuclear import and export signals}, Mol. Cell. Biol. {\bf 21}, 5992 (2001).

\bibitem{itgrn} E Hern\'andez-Lemus, C Rangel-Escare\~no, {\it The role of information theory in gene regulatory network inference}, In: Information theory: New research, Eds. P Deloumeaux, J D Gorzalka, Mathematics Research Developments Series, Nova Publishing (2011).

\raggedbottom
\pagebreak

\bibitem{casm} E Hern\'andez-Lemus, J M Siqueiros-Garc\'ia, {\it Information theoretical methods for complex network structure reconstruction}, Complex Adap. Syst. Mod. {\bf 1}, 8 (2013).

\bibitem{motevo} P Arnold, I Erb, M Pachkov, N Molina, E van Nimwegen, {\it MotEvo: Integrated Bayesian probabilistic methods for inferring regulatory sites and motifs on multiple alignments of DNA sequences}, Bioinformatics {\bf 28}, 487 (2012).


\bibitem{fisher} R A Fisher, {\it Test of significance in harmonic analysis}, P. Roy. Soc. A {\bf 125}, 54 (1929).

\bibitem{rmfchaos} E Hern\'andez-Lemus, K Baca-L\'opez, {\it Bursting and synchronization in gene regulatory dynamics}, Rev. Mex. Fis. {\bf S 58}, 63 (2012).

\end{thebibliography}
\end{document}